\documentclass[10pt]{article}
\usepackage{amsmath,amsfonts,amssymb}
\usepackage{amscd}
\usepackage{amsthm}
\usepackage{multirow}
\usepackage{color}
\usepackage{latexsym}
\usepackage{a4wide}
\usepackage{graphicx}
\usepackage{enumerate}
\usepackage{hyperref}
\usepackage{array}
\theoremstyle{definition}     \newtheorem{remark}{Remark}
\theoremstyle{definition}          \newtheorem{theorem}{Theorem}
\theoremstyle{definition}     
\theoremstyle{definition}          
\theoremstyle{definition}     
\theoremstyle{definition}     
\theoremstyle{definition}     \newtheorem{definition}{Definition}
\theoremstyle{definition}          \newtheorem{proposition}{Proposition}
\theoremstyle{definition}          
\theoremstyle{definition}          
\theoremstyle{definition}          \newtheorem{axiom}{Axiom}

\numberwithin{equation}{section}
\def\Q{\mathcal Q}
\def\P{\mathcal P}
\def\theta#1{\Theta^{\Phi}_{v}(#1)}
\def\thetai#1#2{\Theta^{\Phi}_v(#1,#2)}

\def\thet#1{\Theta^{\Phi^{Sh}}_{v}(#1)}

\def\thetaT#1{\Theta^{\Phi}_{v,T}(#1)}
\def\thetak#1{\Theta^{\Phi}_{v,K_{1}}(#1)}
\def\thetas#1{\Theta^{\Phi^{Sh}}_{v,K_{1}}(#1)}

\begin{document}
\title{{The Expected Shapley value on a class of probabilistic games}}
	\author{Surajit Borkotokey\footnote{Department of Mathematics, Dibrugarh University, India-786004, Email: sborkotokey@dibru.ac.in}~~~~~~Sujata Goala\footnote{Department of Mathematics, Dibrugarh University, India-786004, Email: sujatagoala10@gmail.com}\footnote{Department of Mathematics, Gargaon College, India-785685}~~~~~~~~Rajnish Kumar \footnote{Queen's Management School, Queen's University Belfast,BT9 5EE, UK, E-mail: rajnish.kumar@qub.ac.uk}}
\date{}
\maketitle
\begin{abstract}
\noindent 
We study a class of probabilistic cooperative games which can be treated as an extension of the classical cooperative games with transferable utilities. The coalitions have an exogenous probability of being realized. This probability distribution is known beforehand and the distribution of the expected worth needs to be done before the realization of the state. We obtain a value for this class of games and present three characterizations of this value using natural extensions of the axioms used in the seminal axiomatization of the Shapley value. The value, which we call the Expected Shapley value, allocates the players their expected worth with respect to a probability distribution.

\end{abstract}
\section{Introduction}\label{sec:1}

The classical cooperative games with transferable utilities (TU games) deal with situations where players make coalitions to generate worth. It is assumed that all the players join together to form the grand coalition and a suitable allocation function (we call it a value) determines the share of each player in the worth of this grand coalition. The Shapley value \cite{shapley} is perhaps the most popular allocation function in the literature on TU games. This setup can be termed as the deterministic setup of a TU game. Nevertheless, in reality, the coalition formation process is affected to a great extent by issues relating to players' compatibility among themselves due to viz., their socio-economic, political, or ideological differences. This leads to the definition of a probabilistic cooperative games with transferable utilities (probabilistic TU game) \cite{koster,laruelle} that associates each TU game with a probability of coalition formation. One can think of as an example of this framework to be a sports team, say a cricket team, which has extra players. The extra players are used in the scenario where some player(s) has to drop out due to some reason. Thus the exact composition of the team is not known beforehand. However, every compositions' worth (chance of winning) is known based on the history of the players and the teams. 

The probabilities so assigned to the coalitions reflect the overall level of compatibility among the players in a coalition. Thus, all the coalitions including the grand coalition in a probabilistic TU game are realized with some degree of uncertainty prescribed by an \textit{ex-ante} probability distribution. This prompts the players to form both occasional and long-term relationships, see \cite{ghintan} for a similar argument in the case of a special class of TU games called communication situations due to~\cite{myerson}.   

The procedure of associating probabilities to a TU game is studied in the literature following two approaches. The first approach assumes that the players in a coalition statistically do not depend on each other in forming a coalition and therefore, each player is endowed with an independent probability of joining a coalition. The probability of realizing the subsequent coalition is then obtained using standard methods such as multilinear extension, first proposed in \cite{owen} and subsequently found in the works of \cite{carreras_a,carreras_b} who introduce the multinomial probabilistic value for such games. On the contrary, in \cite{koster,laruelle,vladimir}, it is argued that players depend on each other in deciding whether to join or leave a coalition and therefore, associating independent probabilities with each player can lead to undesirable results under their correlated behavior. Thus, this second approach, as taken in \cite{koster,laruelle,vladimir} is based on the assumption that players do not behave in a silo in deciding to join or leave a coalition, and therefore, the probability distribution is considered as a primitive of the collective decision situation characteristic of a given TU game. In \cite{vladimir}, in addition to the probability of coalition formation, a fourth parameter representing the relationship between two players in terms of their cooperation abilities is taken into account. Our current work follows \cite{koster,laruelle} as we do not consider the fourth parameter proposed in \cite{vladimir}, and therefore, from now onwards, we stick to only these two works. In \cite{laruelle}, a value called the Expected Marginal Contribution (ex-ante and interim), and in \cite{koster} the Prediction value for the class of probabilistic TU games are proposed. Both these values have their origin in the class of probabilistic values first proposed in \cite{weber}. The Expected Marginal contribution due to \cite{laruelle} is interpreted as an assessment of every player's marginal contribution based on the given probability distribution over the coalitions. On the other hand, the Prediction value of a player due to \cite{koster} can be treated as the marginal expectation of a player as it takes the difference of her expectations from the coalitions with respect to the conditional probabilities prescribed by the given probability distribution.

The Expected Shapley value, which we define in our model is the expectation of the Shapley values over all the possible coalitions of the player set with respect to an exogenously given probability distribution. Unlike the Expected Marginal contribution~\cite{laruelle} and the Prediction value~\cite{koster}, ours is a natural extension of the deterministic setup to its probabilistic counterpart. Here, we assume that all the coalitions have some probability to be formed and therefore, the players have some probability to receive payoffs from each coalition under the Shapley value as if it is the final player set that can be formed. We show that most of the properties of the Shapley value for TU games can be extended to this setup. We present three characterizations of the Expected Shapley value typical of its counterpart in TU games, but atypical of the values in \cite{koster,laruelle}. Similar descriptions in the restricted setup of communication situations due to~\cite{myerson} and network games due to~\cite{jw} can be found in the works of \cite{borkotokey_mss,calvo,ghintan,gomez} respectively.

The rest of the paper proceeds as follows. In section \ref{sec:2}, we briefly mention the preliminary ideas relevant to the development of the paper. In section~\ref{sec:3} we study the class of probabilistic TU games and propose the Expected Shapley value for this class. Section~\ref{sec:4} presents the three characterizations of the Expected {Shapley} value. Finally section~\ref{sec:5} concludes.
\section{Preliminary}\label{sec:2}
Let $\aleph$ be  the set of all non-empty and finite subsets of a countably infinite set, we call this the universe of all players. For each $N \in \aleph$, let $2^N$ denote the power set of $N$. The members of $2^N$ are called coalitions and $N$, the largest among them is called the grand coalition. To simplify the notations we use $S \cup i$, $S \setminus i$ etc., instead of $S \cup \{i\}$, $S \setminus \{i\}$ etc. { We use the notation viz., $|S|,|T|$ etc., to denote the size of coalitions $S, T$ etc.} A cooperative game with transferable utilities (TU-game) is a pair $(N, v)$ with $N \in \aleph$ and a coalition function $v: 2^{N} \mapsto  \mathbb{R},$ such that $v(\emptyset) = 0$. The real number $v(S)$ represents the worth of coalition $S \subseteq N$. We also call it the worth generated by $S$. The set of all TU-games with player set $N$ is denoted by $G(N)$ and the set of all TU games with variable players set $N \in \aleph$ is denoted by $G$. If there is no ambiguity with the choice of $N$, we denote the TU game $(N,v)$ only by $v$. 
The restriction of $(N,v)$ to a player set $S \subseteq N$ is denoted by $(S,v)$. The identity game $e_{T} \in G(N)$ is defined as 
\begin{equation}\label{eq:iden}
e_{T}(S)=\left\lbrace\begin{array}{cc}
  1   & \text{if}~  S=T \\
  0   & {\text{otherwise}}
\end{array}\right. 
\end{equation}
and the unanimity  game $u_{T} \in G(N)$ is defined as, 
\begin{equation}\label{eq:unan}
u_{T}(S)=\left\lbrace\begin{array}{cc}
  1   & \text{if}~  T \subseteq S \\
  0   & \text{otherwise}
\end{array}\right.
\end{equation} 
The class of unanimity games $\{u_T : T \subseteq N,\;T \ne \emptyset\}$ and the class of identity games $\{e_T: T \subseteq N, \; T \ne \emptyset\}$ are bases for the game space $G(N)$. 
The null game $(N,v_0)$ is given by, $v_0(S) = 0$ for all $ S \subseteq N$. \\
\noindent Since $G(N)$ is a linear space and $\{u_T : T \subseteq N,\;T \ne \emptyset\}$ is a basis for $G(N)$, every $v\in G(N)$ can be expressed uniquely as a linear combination of these basis vectors as follows:
\begin{equation}\label{eq:har}
v = \sum_{\emptyset \ne S \subseteq N} \Delta_{v}(S)u_S,
\end{equation}
where the term $\Delta_{v}(S)$ is called the Harsanyi dividend~\cite{harsanyi} and is given for all $S \subseteq N$ by,
\begin{equation}\label{eq:dividend}
\Delta_{v}(S)= \left\lbrace \begin{array}{cc}
 0 & \text{if}~ S=\emptyset \\
 v(S)- \sum_{R \subsetneq S} \Delta_{v}(R) & \text{otherwise.}
\end{array}\right.
\end{equation}
An alternative expression of the Harsanyi dividend is given by the following formula.
\begin{equation}\label{eq:hd2}
\Delta_{v}(S)= \sum_{T\subseteq S} (-1)^{|S|-|T|} v(T).
\end{equation}
The marginal contribution of a player $i \in N$ from a coalition $S \subseteq N$ such that $ i \in S$ with respect to a TU game $v \in G(	N)$ is given by 
\begin{equation}\label{eq:mc}
\delta_{v}^{i}(S)=v(S)-v(S\setminus i).
\end{equation} 
\noindent A solution to $G$ is a function defined on $G$ that assigns each TU game a vector of real numbers determined by the size of $N \in \aleph$. An intuitive assumption in this framework is that for each $N \in \aleph$, called the grand coalition, the $|N|$-vector given by a solution is usually a distribution of the worth of this grand coalition. The single point solutions are called values. The most popular and transverse value is the Shapley value~\cite{shapley}, defined in terms of Harsanyi dividends~\cite{harsanyi} as follows,
\begin{equation}\label{eq:Shapley1}
\Phi^{Sh}_{i} (N,v)=\sum_{ S \subseteq N : i \in S}\frac{\Delta_{v}(S)}{|S|},~~\forall~i \in N.
\end{equation} 
Alternatively, the Shapley value is also expressed as 
\begin{equation}\label{eq:shapley2}
\Phi^{Sh}_{i} (N,v)=\sum_{ S \subseteq N \setminus i} \frac{(|N|-|S|-1)!|S|!}{|N|!}\delta_v^i(S).
\end{equation}
\noindent The Shapley value is characterized, among others by Shapley~\cite{shapley}, Myerson~\cite{myerson} and Hart and {Mas-Colell}~\cite{Hart} using three sets of axioms. We mention briefly these characterizations as we will use them when we define the Expected Shapley value at a latter stage. As an apriori requirement, we give the following definitions.
\begin{definition}
Players $i,j\in N$ are called symmetric in $(N, v)\in G$ if for all $S \subseteq N\setminus \{i,j\}$, $v(S\cup i) = v(S \cup j)$.
\end{definition}
\begin{definition}
Player $i\in N$ is a null player in $(N, v)\in G$ if for all $S \subseteq N\setminus i$, $v(S\cup i) = v(S)$. We call a player productive to $(N, v)$ if $v(S\cup i)> v(S)$ for at least one $S \subseteq N\setminus i$. Thus, a null player is non-productive to the game $(N, v)$.
\end{definition}
\noindent The corresponding axioms are given as follows.
\begin{axiom}[Null player property (NP)]
A value $\Phi$ on $G$ satisfies the Null player property if $\Phi_i(N,v) = 0$ for each null player $i\in N$.
\end{axiom}
\begin{axiom}[Symmetry (SYM)]
A value $\Phi$ on $G$ satisfies Symmetry if $\Phi_i(N, v) = \Phi_j(N, v)$ for each pair of symmetric players $i, j\in N$.
\end{axiom}
\begin{axiom}[Efficiency (E)]
A value $\Phi$ on $G$ satisfies Efficiency (E) i.e., $\displaystyle \sum_{i\in N} \Phi_i(N, v) = v(N)$.
\end{axiom}
\begin{axiom}[Linearity (LIN)]
A value $\Phi$ on $G$ is Linear if for $(N, v),\; (N, w) \in G$, and $\alpha, \beta\in \mathbb R$, one must have 
$$\Phi(N, \alpha v + \beta w)= \alpha \Phi(N, v) + \beta \Phi(N, w).$$
$\Phi$ is Additive (ADD) if the above conditions holds only for $\alpha=\beta = 1$.
\end{axiom}
\noindent Shapley~\cite{shapley} uses E, SYM, NP and LIN (ADD) for the characterization of the Shapley value. On the other hand, Myerson~\cite{myerson} characterizes it using only two axioms: E and a new axiom, namely, Balanced Contribution (BC). The Balanced Contribution requires that the amount of one player's gains or losses is equal to the other player's gains or losses when the two players reverse their roles of leaving the game. Thus, formally we have:
\begin{axiom}[Balanced Contribution (BC)]
For all $N \in \aleph$, $(N, v) \in G$ and $i,j\in N$, we have
\begin{equation}
\Phi_i(N,v)-\Phi_i(N\setminus j, v) = \Phi_j(N,v)-\Phi_j(N\setminus i, v)
\end{equation}
where, the game $(N\setminus k, v)$ is the restriction of $v$ on $N\setminus k$ for $k \in \{i,j\}$.
\end{axiom}
\noindent The BC axiom has been instrumental in characterizing the Shapley value on restricted TU games such as communication situations~\cite{myerson78}, network games~\cite{jw} and their probabilistic counterparts, see for example~\cite{borkotokey_mss,calvo,ghintan,gomez} etc. Following these approaches, we will also characterize the Expected Shapley value at a latter stage. An alternative approach is taken by Hart and {Mas-Colell}~\cite{Hart} who obtain the potential function for TU games and characterize the Shapley value using the axioms: Consistency and Standard for two person games. Given the class $G$ of all games over all player sets $N \in \aleph$, and a function $P : G \mapsto \mathbb R$ which associates a real number $P(N, v)$ to every game $(N, v)$, the marginal contribution of player $i \in N$ in $(N, v)$ with respect to $P$ is given by 
\begin{equation}\label{eq:potential_mc}
D^i P(N, v)= P(N, v) - P(N\setminus i, v) 
\end{equation}
where $(N\setminus i, v)$ is the restriction of $(N, v)$ to $N\setminus i$.
\begin{definition}
A function $P: G \mapsto \mathbb R$ satisfying $P(\emptyset, v)=0$ is called a potential function if it satisfies the following condition:
\begin{equation}\label{eq:potential}
\sum_{i \in N}D^i P(N, v)= v(N)\;\;\forall\;(N,v) \in G. 
\end{equation}
\end{definition}
\noindent The following theorem from~\cite{Hart} gives the existence and uniqueness of a potential function and its connection to the Shapley value.
\begin{theorem}[\cite{Hart}, page 591] There exists a unique potential function $P$. For every game $(N, v)$, the resulting payoff vector $(D^i P(N, v))_{i\in N}$ of the derivatives coincides with the Shapley value of the game. Moreover, the potential of any game $(N, v)$ is uniquely determined by Eq.(\ref{eq:potential}) applied only to the game and its subgames (i.e., to $(S, v)$ for all $S \subset N$). 
\end{theorem}
\noindent In \cite{Hart}, the two axioms used to characterize the Shapley value are Consistency (CON) and Standard for two person games (STPG). Before describing these axioms, we define the following. 
\begin{definition}
Let $\Phi$ be a function defined on $G$, $(N,v)$ a game and $T \subset N$ a coalition. The reduced game $(T, v^\Phi_T)$ is defined as
\begin{equation}\label{eq:reduced}
v^\Phi_T (S) = v( S \cup T^c) - \sum_{i \in T^c} \Phi_i(S \cup T^c, v)\;\;\;\forall S \subset T,
\end{equation}
and $v^\Phi_T (\emptyset)=0 $.\\

\end{definition}
\noindent In the following, we obtain the {new} reduced game using a notion called value dividend. This will be used to define the probabilistic version of the reduced game later. We will also show that through this new approach, the inter connection between the reduced game of a classical TU game and its counterpart under the probabilistic framework can be presented in an intuitive manner. This way, we pragmatically deviate from the existing literature of a reduced game given by Eq.(\ref{eq:reduced}) due to \cite{Hart} and {use} the following definition of a value dividend {due to ~\cite{Besner-2020}}.
\begin{definition}
For  each $(N,v) \in G,~S \subseteq N$ and  a  value $\Phi$ on $G$, the value dividends of any player $i\in S$ from coalition $S$ with respect to $\Phi$ denoted by $\thetai i S$ are defined inductively by,   
\begin{equation}\label{eq:value-dividend}
\thetai i S =\left\lbrace \begin{array}{cc}
 \Phi_{i}(\{i\},v) & \text{if}~ S=\{i\} \\ 
 \Phi_{i}(S,v)-\displaystyle\sum_{K \subsetneq S: i \in K} \thetai i K & \text{otherwise}
\end{array}\right.
\end{equation}  
{The value dividend of the coalition $S$ corresponding to the value $\Phi$ is then given by,
\begin{equation}
\theta S = \displaystyle \sum_{i\in S} \thetai i S, ~\text{and}~ \theta \emptyset=0
\end{equation}
The $T$ reduced value dividend of the coalition $K$ corresponding to the value $\Phi$ is then given by
\begin{equation}\label{eq:reduced:vd}
\thetaT K = \displaystyle \sum_{i\in K} \thetai i K - \sum_{i\in K \setminus T} \thetai i K=\displaystyle \sum_{i\in T} \thetai i K~~~\text{for all $T \subseteq K$, and $\thetaT \emptyset = 0$}
\end{equation}}
\end{definition}
\begin{remark}\label{rem:1}
\noindent {In particular for $K=T$, $\thetaT T =\theta T$. Also by remark 3.2 of ~\cite{Besner-2020}, the value dividend for a coalition is identical to the Harsanyi dividend whenever $\Phi$ is efficient. Therefore, it follows that $\theta T =\Delta_{v}(T)$ for $\Phi \equiv \Phi^{Sh}$.} { Consequently we also have, $v(S)=\sum_{T \subseteq S}\Delta_{v}(T)=\sum_{T \subseteq S} \thet T$. In our proposed model in this current study, we pay our attention only to the Shapley values and their convex combinations over probability distributions, therefore, we do not distinguish between the value dividend and the Harsanyi dividend in our framework, rather we treat them as alternatives.}
\end{remark} 

\noindent Using the definition of the  reduced   value dividend $\thetaT K$  for a coalition $K$ given by Eq.(\ref{eq:reduced:vd}) we define a reduced game ${v^{*}_{T}}^{\Phi}$ as follows:
\begin{eqnarray}\label{eq:vdreduced}
{v^{*}_{T}}^{\Phi}(S)&=&\sum_{\substack{ K_{1} \subseteq S; K_{2} \subseteq  T^{c}\\ K_1  \neq \emptyset }}\thetak   {K_{1} \cup K_{2}},~~\forall S \subseteq T. 
\end{eqnarray}
and ${v^{*}_{T}}^{\Phi}(\emptyset)=0$.\\
Note that for the class of efficient values Eq.\eqref{eq:vdreduced} is identical to Eq.\eqref{eq:reduced}. { As already mentioned, we confine our study only to the properties of the Shapley values and their convex combinations, we treat the reduced game defined by Eq.(\ref{eq:reduced}) equivalent to the reduced game defined in Eq.(\ref{eq:vdreduced}) and use the same notation viz., $v_T^{\Phi}$ for both.}

\begin{remark}\label{rem:2}
Observe that the reduced game on a coalition $T$ with respect to a value is defined by rescaling the worth of the coalitions so that the players outside $T$ get their payoffs according to this value and leave the game. The reduced game in value dividend form given by Eq.(\ref{eq:vdreduced}) is therefore, a rescaling of the worth given by the original game $(N, v)$ determined by the value dividend $\Theta^\Phi_v$ so that the players outside $T$ leave the game with their payoffs according to $\Phi$. We make use of this idea in section~\ref{sec:4} again. 
The reduced game with respect to the Shapley value is given by the following.
{\begin{eqnarray}\label{eq:vdreducedSh}
v_{T}^{\Phi^{Sh}}(S)&=&\sum_{\substack{ K_{1} \subseteq S; K_{2} \subseteq  T^{c}\\ K_{1} \neq \emptyset }}\thetas   {K_{1} \cup K_{2}},~~\forall S \subseteq T.
\end{eqnarray}}
and $v_{T}^{\Phi^{Sh}}(\emptyset)=0$. \\
It follows from Eq.(\ref{eq:value-dividend}) that $\thetai  i S =\frac{\Delta_{v}(S)}{|S|}=\thetai j S= \frac{\Theta^{\Phi^{Sh}}_v(S)}{|S|}$, for all $i,j \in S$. In view of this relationship between the Harsanyi dividend and the value dividend with respect to the Shapley value, Eq.(\ref{eq:vdreducedSh}) can also be put in the following form:
\begin{eqnarray}\label{eq:vdredSh_1}
v_{T}^{\Phi^{Sh}}(S)
&=& \sum_{\substack{K_{1} \subseteq S; K_{2} \subseteq T^{c}\\K_{1}  \neq \emptyset}}|K_{1}|\cdot\frac{\Theta^{\Phi^{Sh}}_v(K_{1}\cup K_{2})}{|K_{1}|+|K_{2}|},~~\forall ~S \subseteq T.  
\end{eqnarray}
and $v_{T}^{\Phi^{Sh}}(\emptyset)=0$.\\
Moreover, under the new notation of the value dividend, the Shapley value $\Phi^{Sh}$ has the following equivalent form:
\begin{equation}\label{eq:vdsh}
\Phi_i^{Sh}(N, v)= \sum_{ S \subseteq N : i \in S}\frac{\Delta_{v}(S)}{|S|}=\sum_{ S \subseteq N : i \in S} \frac{\Theta^{\Phi^{Sh}}_v(S)}{|S|},~~\forall~i \in N.
\end{equation}
\end{remark}
\noindent The following two axioms are due to \cite{Hart}. 
\begin{axiom}[Consistency (CON)]  
A value $\Phi$ is consistent if for every game $(N, v) \in G$ and $T \subset N$, we have
\begin{equation}\label{eq:consistency}
{\Phi_j(T, v^\Phi_T) = \Phi_j(N, v)\;\;\forall j\in T.}
\end{equation}
\end{axiom}
\begin{axiom}[Standard for two person games (STPG)]
A value $\Phi$ is standard for two person games if 
\begin{equation}\label{eq:standard}
\Phi_i(\{i,j\}, v) = v(i)+ \frac{1}{2}\{v(\{i,j\}) - v(i) -v(j)\}\;\;\forall i \ne j\;\forall v\in G.
\end{equation}
\end{axiom}
{\begin{axiom}[$*$-Consistency ($*$-CON)]  
A value $\Phi$ is $*$-Consistent if for every game $(N,v) \in G$ and $T \subset N$, we have
\begin{equation}\label{eq:*consistency}
\Phi_j(T,{v^{*}}^\Phi_T) = \Phi_j(N,v)\;\;\forall j \in T.
\end{equation}
\end{axiom}
\begin{remark}
Similar to the observations made in remark~\ref{rem:1} and remark~\ref{rem:2}, we can show that for the class of efficient values, $*$-CON is equivalent to CON. Since the Shapley value is efficient we treat the two properties for the Shapley value as equivalent and use CON for both ``$*$-CON" and ``CON".    
\end{remark}}
\noindent The three characterizations of the Shapley value due to \cite{Hart,myerson,shapley} as mentioned above are formally given in the following theorem. We will make use of these results for the characterization of the Expected Shapley value at a latter stage.
\begin{theorem}
Let $\Phi$ be a value on $G$. Then the following statements are equivalent:
\begin{enumerate}[~~(a)]
\item $\Phi$ is the Shapley value on $G$.
\item $\Phi$ satisfies E, SYM, NP and LIN (ADD).
\item $\Phi$ satisfies E and BC.
\item $\Phi$ satisfies CON and is STPG.
\end{enumerate}
\end{theorem}
\noindent In the following section we present the notion of a probabilistic TU game.
\section{Probabilistic TU games}\label{sec:3}
Following our discussion in section~\ref{sec:1} and the works of \cite{koster,laruelle}, now we assume that the formation of the coalitions is realized with a probability distribution over the set $2^N$ of coalitions of $N$. Thus, for each $S \subseteq N$, we associate a probability measure $p^N : 2^N \mapsto [0,1]$ such that the set $\{p^N(S)| S \subseteq N,\;\sum_{S \subseteq N} p^N(S)=1\}$ forms a probability distribution over $2^N$. With an abuse of notation, we call the probability measure $p^N$ a coalition formation probability distribution (CFPD in short). Denote the set of all such distributions by $ \mathbb{P}^{N}$. Thus formally, we have 
$$ \mathbb{P}^{N} = \Big\{p^{N}: 2^N \mapsto [0,1]|\sum_{S\subseteq N} p^{N}(S) =1\Big\}.$$
Let $\mathbb{P}$ denote the set of all CFPDs over a variable player set $N \in \aleph$ i.e., $\mathbb{P}=\bigcup\{\mathbb{P}^{N}:~N \in \aleph \}$.

The set containing all coalitions $S \subseteq N$ for which $p^{N}(S) > 0$ is called the support of $p^{N}$. It is denoted by $N(p^{N})$. Thus, formally we have $$N(p^{N})=\Big\{S \subseteq N: p^{N}(S) > 0\Big\}.$$
Clearly, $N(p^{N}) \subseteq 2^{N}$. Let us now define the restriction of a CFPD over $N$ to a subset $M$ of $N$. We assume that under this restriction, the probabilities of formation of the coalitions outside $M$ are all zero and therefore, the restricted probability distribution can be treated as a conditional probability distribution on $M$ through rescaling the probabilities in $p^N$ by ruling out the contributions of the players outside $M$. A similar formulation can be found in~\cite{gomez} in communication situations. Formally the restriction of the CFPD is defined as follows:
\begin{definition}
Let $p^{N} \in \mathbb{P}^{N}$ be a CFPD and  $M \subseteq N$. Then the restriction of $p^{N}$ to $M$ is the modified CFPD $p^{N}_{M} \in \mathbb{P}^{N}$ defined by,  
\begin{equation}\label{eq:modifiedp}
p^{N}_{M}(S)=\left\lbrace \begin{array}{cc}
\displaystyle\sum_{T\subseteq N \setminus M} p^{N}(S \cup T) ,& \forall S\subseteq M\\
0 & \text{otherwise.}
\end{array} \right.
\end{equation}
\end{definition}
In particular, denote  $p^{N}_{N \setminus i} \in \mathbb{P}^{N}$ by $p^{N}_{-i}\in \mathbb{P}^{N}$ so that, 
\begin{eqnarray}
p^{N}_{-i}(S)=p^{N}_{N \setminus i}(S)&=&\left\lbrace \begin{array}{cc}
\displaystyle\sum_{T\subseteq N \setminus (N \setminus i)} p^{N}(S \cup T) ,& \forall S \subseteq N \setminus i\\
0 & \text{otherwise}.
\end{array} \right.\nonumber\\
&=&\left\lbrace \begin{array}{cc}
\displaystyle\sum_{T\subseteq \{ i\}} p^{N}(S \cup T) ,& \forall S \subseteq N \setminus i \\
0 & \text{otherwise}.
\end{array} \right.\nonumber\\
&=&\left\lbrace \begin{array}{cc}
 p^{N}(S \cup i)+ p^{N}(S),& \forall S \subseteq N \setminus i \\
0 & \text{otherwise}.\label{eq:p_i}
\end{array} \right.
\end{eqnarray}
Note that the restricted probability distribution given by Eq.(\ref{eq:p_i}) is also found in \cite{koster,laruelle}, in the name of conditional probability distribution such that player $i$'s contributions to the game is ignored, however, here we consider this as a special case of the more general version given by Eq.(\ref{eq:modifiedp}). Following the same line of arguments, we interpret the probability distribution $p^N_M$ given by Eq.(\ref{eq:modifiedp}) as the conditional probability distribution over the player set $M$ such that the players in $N \setminus M$ are ignored. Now, we define a probabilistic TU game as follows. 
\begin{definition}
A probabilistic TU game is a triple $(N, v, p^{N}) \in G \times \mathbb{P}$ consisting of a TU game $(N, v)$ that represents the potential worth generated from the coalitions of $N$ and a coalition formation probability distribution $p^{N}$ that describes the probability with which the coalitions are formed from the player set $N \in \aleph$.
\end{definition}
\noindent The expected worth, that is generated by the probabilistic TU game $(N, v, p^N) \in G \times \mathbb P$ is given by
\begin{equation}\label{eq:exp}
\mathbb{E}(N,v,p^{N})=\sum_{S \in N(p^{N})}p^{N}(S)v(S).
\end{equation}
\noindent Note that in \cite{koster}, the expected worth from a probabilistic TU game $(N, v, p)$ is denoted by $\mathbb E_p[v(S)]$ where the coalition $S$ seems to be redundant. Therefore, we depart from this notation and use the one as in Eq.(\ref{eq:exp}) instead. Unlike in the case of TU games where the worth of the grand coalition is assumed to be allocated among the players by a suitable value function, here the expected worth is allocated among the players under a suitable value function over $G \times \mathbb P$. This we present as an axiom in our characterization, namely the Expected Efficiency axiom. The next proposition gives us an expression of the expected worth of a probabilistic TU game under a restricted probability distribution.
\begin{proposition}\label{prop:1}
\rm Let $(N,v,p^{N}) \in G\times \mathbb{P}$ be a probabilistic TU game then, 
$$\mathbb{E}(N, v, p^{N}_{-i})=\mathbb{E}(N,v,p^{N})-\sum_{S\subseteq N: i \in S}p^{N}(S)\delta^{v}_{i}(S), $$
where $\delta_i^v(S)$ is the marginal contribution of player $i \in N$ in $S \subseteq N$ with respect to $(N,v)\in G$ given by Eq.(\ref{eq:mc}). 
\end{proposition}
\begin{proof}
Using the expression of $p^{N}_{-i}$ given in Eq.(\ref{eq:p_i}), we get from Eq.(\ref{eq:exp})
\begin{eqnarray*}
\mathbb E(N, v, p^N_{-i})&=& \sum_{S\subseteq N }p^{N}_{-i}(S)v(S)  \\
&=& \sum_{S\subseteq N \setminus i}\Big\{\sum_{K \subseteq \{ i\}}p^{N}(S \cup K)\Big \}v(S)\\
&=& \sum_{S\subseteq N \setminus i} \Big\{p^{N}(S)+p^{N}(S \cup i)\Big\}v(S)\\
&=& \sum_{S\subseteq N \setminus i}p^{N}(S)v(S)+\sum_{S\subseteq N \setminus i}p^{N}(S \cup i)v(S)\\
&=& \sum_{S\subseteq N \setminus i}p^{N}(S)v(S)+ \sum_{S\subseteq N : i \in S}p^{N}(S)v(S)\\
&&-\sum_{S\subseteq N : i \in S}p^{N}(S)v(S)+ \sum_{S\subseteq N : i \in S}p^{N}(S)v(S \setminus i)\\
&=& \sum_{S\subseteq N}p^{N}(S)v(S)-\Big\{\sum_{S\subseteq N: i \in S}p^{N}(S)\delta_{i}^{v}(S)\Big\}\\
&=& \mathbb{E}(N,v,p^{N})-\Big\{\sum_{S\subseteq N: i \in S}p^{N}(S)\delta_{i}^{v}(S)\Big\}.
\end{eqnarray*}
This completes the proof.
\end{proof}
\begin{remark}
In view of proposition~\ref{prop:1} and Eq.(\ref{eq:mc}), we call the expression $\mathbb{E}(N,v,p^{N})-\mathbb{E}(N,v,p^{N}_{-i})$  the expected marginal contribution of player $i\in N$ from the probabilistic game $(N, v, p^N)$. Recall that in \cite{laruelle} also, the term expected marginal contribution is used to denote the probabilistic value defined there. Similarly, in \cite{koster}, the prediction value is defined to be the difference of the expectations with respect to each player, however, in either of the values, the expectations are not computed with the restricted probability specific to a particular coalition where the contributions of the players outside this coalition should be ignored. In our model, while computing the expectation of the Shapley values at all possible coalitions, we seem to ignore the contributions of the players outside these coalitions by rescaling the probability distribution to each of these coalitions according to Eq.(\ref{eq:modifiedp}). This is justified in the sense that each coalition has a probability to be realized as the grand or final coalition among the players. 
\end{remark}
\noindent In what follows next, we define a value on $G \times \mathbb P$ on the class of probabilistic TU games, see \cite{koster}. We also define the Expected Shapley value in this new framework.
\begin{definition} 
A value for the class of probabilistic TU games $G \times \mathbb P$ is a mapping $\Phi$ that assigns to each  $(N, v, p^{N}) \in G\times \mathbb{P}$ a vector $\Phi(N, v, p^N) \in \mathbb R^{|N|}$. The $i$-th component $\Phi_i(N, v, p^N)$ of the vector denotes the measure of the difference that player $i$ makes for the probabilistic game $(N, v, p^N)$. We call this the payoff to player $i$ following standard game theoretic terminologies.
\end{definition}
\begin{definition}\label{def:psh}
The Expected Shapley value $\Phi^{Exp-Sh}$ defined on $G\times \mathbb{P}$ is given by,
\begin{equation}\label{eq:psh}
\Phi^{Exp-Sh}_i(N,v,p^{N})=\sum_{{S \in N(p^{N}): i \in S}}p^{N}(S)\Phi^{Sh}_i(S,v)
\end{equation}  
Note that $\Phi^{Sh}_i(S,v)=0$ for all $i \in N \setminus S$.
\end{definition}
\noindent It follows from definition \ref{def:psh}~~that the Expected Shapley value given by Eq.(\ref{eq:psh}) is the expectation over all Shapley values on the coalitions of the player set $N \in \aleph$ with respect to the probability distribution $p^N$ and the restricted game  $(S,v)$ for each $S\subseteq N$. 
\begin{remark}
In line with Eq.(\ref{eq:Shapley1}) and hence with Eq.(\ref{eq:vdsh}), the Expected Shapley value can be obtained in Harsanyi and value dividend form as follows:
\begin{eqnarray*}
\Phi_{i}^{Exp-Sh}(N, v,p^{N})&=&\sum_{\substack{{S \in N(p^{N}})\\i\in S}} p^{N}(S)\Phi_{i}^{Sh}(S,v)\\
&=&\sum_{\substack{{S \in N(p^{N}})\\i\in S}} p^{N}(S)\sum_{T \subseteq S:i \in T}\frac{\Delta_{v}(T)}{|T|}\\
&=&\sum_{\substack{{S \in N(p^{N}})\\i\in S}}\frac{\Delta_{v}(S)}{|S|}\Big\{\sum_{K\subseteq N \setminus S}p^{N}(S \cup K)\Big\}\\
&=&\sum_{\substack{{S \in N(p^{N}})\\ i\in S}}\frac{\Delta_{v}(S)}{|S|}p^{N}_{S}(S)\\
&=&\sum_{\substack{{S \in N(p^{N}})\\ i\in S}}{\frac{\Theta_{v}^{\Phi^{Sh}} (S)}{|S|}\cdot p^{N}_{S}(S)}
\end{eqnarray*}
\end{remark}
In the next section we present three characterizations of the Expected Shapley value.
\section{Characterization of the Expected Shapley value}\label{sec:4}
Following Shapley's characterization of the Shapley value for classical TU games given in \cite{shapley}, we provide the first characterization of the Expected Shapley value using the axioms: Expected Efficiency, Expected Null player property, Compatibility and Additivity. The second characterization is done using the axioms Expected Efficiency and Expected Balanced Contribution in line with their counterparts in classical TU games given by Myerson~\cite{myerson}. Finally, we define the potential of a probabilistic TU game and show that the Expected Shapley value is expressible in terms of the potential function. Moreover, we introduce the probabilistic version of the axioms: Consistency and Standard for two person games and give a characterization of the Expected Shapley value in terms of these two axioms.
\subsection{The First Characterization}\label{subsec:4.1}
The first axiom namely, Expected Efficiency is presented as follows.
\begin{axiom}[Expected Efficiency (EE)] A value on $G \times \mathbb P$ is said to satisfy  Expected Efficiency (EE) i.e., $$\sum_{i \in N}\Phi_{i}(N, v,p^{N})=\mathbb{E}(N, v,p^{N}).$$
\end{axiom}
It follows from the axiom EE that, a value that satisfies this axiom allocates the expected worth of the probabilistic TU game among the players. The characterization of the expected marginal contribution in \cite{laruelle} and the Prediction value in \cite{koster} do not use efficiency in general as it is counter intuitive in their models to assume that the payoffs to the players according to these two values should add up to the worth of the grand coalition. However, efficiency (i.e., EE) in our model appears quite naturally as the payoffs to the players, which we can call their expected payoffs should add up to the expected worth generated by the game. For the Expected Null player property, we define a p-Null player as follows.
\begin{definition}
 A player $i \in N$ is said to be a p-Null player in the probabilistic TU game  $(N,v,p^{N})$, if ~$\mathbb{E}(N, v,p^{N})-\mathbb{E}(N, v,p^{N}_{-i})=0$.
 \end{definition}
\begin{remark}
The p-Null player is one whose expected marginal contribution is zero, i.e., she may be productive to the TU game $(N, v)$, however, her contributions are ignored by the probability distribution. On the other hand it is also possible that her contributions are acknowledged by the probability distribution, but the game does not consider her to be productive. Thus, in either case, it is natural to axiom that the p-Null player should get zero payoff from the game. Also observe that, in view of proposition~\ref{prop:1}~ we note that player $i\in N$ is a p-Null player in the probabilistic TU game $(N, v, p^N)$ if~ $\displaystyle\sum_{S \subseteq N: i \in S}p^{N}(S)\delta_{i}^{v}(S)=0$.
\end{remark}
\noindent The corresponding Expected Null player property goes as follows. 
\begin{axiom}[Expected Null Player Property (ENP)]
A value on $G \times \mathbb P$ satisfies the Expected Null Player Property (ENP) namely, $\Phi_{i}(N, v, p^{N})=0$, for each p-Null player $i \in N$.
\end{axiom}
\noindent The next axiom is on ``compatibility" between two players which resembles with the symmetric players in classical TU games. We first define the notion of compatibility under the probabilistic setups and then introduce the corresponding axiom.
\begin{definition}\label{def:comp}
Player $i,j \in N(p^{N})$ are compatible  in $(N, v,p^{N})$ if, 
\begin{equation}\label{eq:comp}
\mathbb{E}(N,v,p^{N})-\mathbb{E}(N,v,p^{N}_{-i})=\mathbb{E}(N, v,p^{N})-\mathbb{E}(N,v,p^{N}_{-j}).
\end{equation} 
\end{definition}
\noindent It follows from definition~\ref{def:comp} that two players are compatible if their expected marginal contributions are equal. 
\begin{axiom}[Compatibility (COM)]
A value $\Phi$ on $G \times \mathbb P$  satisfies Compatibility (COM) i.e., $\Phi_{i}(N, v,p^{N})=\Phi_{j}(N, v,p^{N})$ for a pair of compatible  players $i,j \in N(p^{N})$ in the probabilistic game $(N, v, p^N)$. 
 \end{axiom}
 \begin{axiom}[Additivity (ADD)]
 A value $\Phi$ on $G \times \mathbb P$ satisfies Additivity (ADD)  if given $(N,v),\;(N,w)\in G$ and $p^{N} \in \mathbb{P}$ we have, $$\Phi(N, v+w,p^{N})=\Phi(N, v,p^{N})+\Phi(N, w,p^{N}).$$
 \end{axiom}
\begin{theorem}\label{them:ESLN}
A value $\Phi:G\times \mathbb{P} \to \mathbb{R}^{|N|}$ satisfies EE, ENP, COM and ADD if and only if $\Phi\equiv \Phi^{Exp-Sh}$.
\end{theorem}
\begin{proof}
First, we show that the Expected Shapley value satisfies the given axioms.\\
The Expected Shapley value inherits EE from axiom E of the  Shapley value of the corresponding coalitional TU game $(N, v)$. Thus, we have
\begin{eqnarray*}
\sum_{i \in N}\Phi_{i}^{Exp-Sh}(N,v,p^{N})&=&\sum_{i \in {S \in N(p^{N})}}p^{N}(S)\sum_{i \in S}\Phi_{i}^{Sh}(S,v)\\
&=&\sum_{S \in N(p^{N})}p^{N}(S)v(S)\\
&=&\mathbb{E}(N,v,p^{N}).
\end{eqnarray*}
\noindent Note that, $i$ is a p-Null player in $(N, v,p^{N})$ if 
\begin{eqnarray*}
0 &=& \mathbb{E}(N,v,p^{N})-\mathbb{E}(N,v, p^{N}_{-i})\\
  &=&\sum_{{S \in N(p^{N})}:i \in S}p^{N}(S)\Big\{v(S)-v(S \setminus i)\Big\}\\
  &=&\sum_{\substack {{S \in N(p^{N})}\\i \in S}}p^{N}(S)\Big\{v(S)-v(S \setminus i)\Big\}.
\end{eqnarray*}
Since $p^{N}(S)>0$ for all ${S \in N(p^{N})}: i \in S$, we have, $v(S)-v(S \setminus i)=0, \forall {S \in N(p^{N})}:i \in S$, i.e., for all ${S \in N(p^{N})}: i \in S$, $i$ is a Null player in $(S,v) \in G$. Thus, by NP of the Shapley value, $\Phi^{Sh}_{i}(S,v)=0, \forall {S \in N(p^{N})}:i \in S$.\\
Therefore, 
$$\Phi_{i}^{Exp-Sh}(N, v,p^{N})=\displaystyle\sum_{{S \in N(p^{N})}:i \in S}p^{N}(S)\Phi_{i}^{Sh}(S,v)=0, \;\textrm{for each p-Null player $i$ in  $(N, v,p^{N})$.}$$ 
Let players $i$ and $j$ be compatible players in the probabilistic TU game $(N, v, p^N)$. Therefore, we have, 
$$\mathbb{E}(N,v,p^{N})-\mathbb{E}(N,v,p^{N}_{-i})=\mathbb{E}(N,v,p^{N})-\mathbb{E}(N,v,p^{N}_{-j}).$$
Then, using the fact that adding terms for $S \in N(p^N)$ such that $j\in S$ and $i\in S$ on both sides of the following expressions does not change the equality and also the fact that $p^{N}(S) > 0, ~~\forall S \subseteq N(p^{N})\setminus \{i,j\}$, we get,
\begin{eqnarray*}
\sum_{ {S \in N(p^{N})}:i \in S}p^{N}(S)\Big\{v(S)-v(S \setminus i)\Big\}=\sum_{ {S \in N(p^{N})}:j \in S}p^{N}(S)\Big\{v(S)-v(S \setminus j)\Big\}~~~~~~~~~~~~\\
\Rightarrow \sum_{{S \in N(p^{N})}: i,j \notin S} p^{N}(S)\Big\{v(S \cup i)-v(S \cup j)\Big\}=0~~~~~~~~~~~~~~~~~~~~~~~\\
~~~~~~~~~~~~~~~~~~~~~~~~~~~~~~~~~~~~~\\
\Rightarrow  v(S \cup i)=v(S \cup j),~~ \forall ~~{\{i,j\} \subsetneq S \in N(p^{N})}.~~~~~~~~~~~~~~~~~~~~~~~~~~~~
\end{eqnarray*} 
Therefore  for any   { $K \in N(p^{N})$}, $i$ and $j$ are also symmetric players in $(K, v) \in G$ where $i, j \in K$. Therefore, by SYM of the Shapley value, $\Phi_{i}^{Sh}(K,v)=\Phi_{j}^{Sh}(K,v)$ for all {$K \in N(p^{N})$} such that $i, j \in K$. It follows that $$\Phi_{i}^{Exp-Sh}(N, v,p^{N})=\Phi_{j}^{Exp-Sh}(N, v,p^{N}).$$ 
For Additivity of the Expected Shapley value, we have the following. Let $(N, v, p^N)$ and $(N, w, p^N)$ be two probabilistic TU games defined on the same player set $N$ and the same coalition formation probability distribution $p^N$. Then,
\begin{eqnarray*}
\Phi_{i}^{Exp-Sh}(N, v+w, p^{N})&=&\sum_{S \in N(p^{N})}p^{N}(S)\Phi_{i}^{Sh}(N(p^{N}),v+w)\\
&=& \sum_{S \in N(p^{N})}p^{N}(S)\Phi_{i}^{Sh}(N(p^{N}),v)+\sum_{S \in N(p^{N})}p^{N}(S)\Phi_{i}^{Sh}(N(p^{N}),w)\\
&=&\Phi_{i}^{Exp-Sh}(N, v,p^{N})+\Phi_{i}^{Exp-Sh}(N, w,p^{N}).
\end{eqnarray*} 
Therefore,  $\Phi^{Exp-Sh}$ satisfies all the four axioms. For the converse part, let $(N,v,p^{N})\in G\times \mathbb{P}$ be a probabilistic TU game and $\Phi$ satisfy the above four axioms. Recall from section~\ref{sec:2} that each $v$ can be uniquely expressed as, $v=\displaystyle\sum_{\emptyset\neq T\subseteq N}c_{T}u_{T},\; c_{T}\in \mathbb{R}$. Then the expected worth of $(N,v,p^{N})$ i.e.,
$\mathbb{E}(N,v,p^{N})$ is given by,\\
\begin{eqnarray*}
\mathbb{E}(N,v,p^{N})&=&\sum_{S\subseteq N}p^{N}(S)v(S)\\
&=&\sum_{S\subseteq N}p^{N}(S)\sum_{T\subseteq N}c_{T}u_{T}(S)\\
&=&\sum_{T\subseteq N}c_{T}(\sum_{S\subseteq N}p^{N}(S)u_{T}(S))\\
&=&\sum_{T \subseteq N}c_{T}\mathbb{E}(N, u_{T},p^{N}).
\end{eqnarray*}
By ADD, $\Phi(N,v,p^{N})=\displaystyle\Phi(N,\sum_{T\subseteq N}c_{T}u_{T},p^{N})=\sum_{T\subseteq N}\Phi(N, c_{T}u_{T},p^{N})$. Thus, it is sufficient to show that $\Phi(N, c_{T}u_{T},p^{N})=\Phi^{Exp-Sh}(N, c_{T}u_{T},p^{N})$ . \\
Without loss of generality we assume that, $c_{T}=1$. Note that in $(N, u_{T},p^{N})$, each $i\in T \subseteq  N(p^{N})$ is compatible with any $j \in T$, and every $i \notin T$ is a p-Null player. This can be easily seen from the following expression:
\begin{equation}\label{eq:mc1}
\mathbb E(N, u_T, p^N) - \mathbb E(N, u_T, p^N_{-i}) = \sum_{{S \in N(p^{N})}: i\in S}p^N(S)\Big\{u_T(S)-u_T(S\setminus i)\Big\}.
\end{equation}
For $i \not\in T$, $u_T(S) = u_T(S\setminus i)$ and therefore, it follows from Eq.(\ref{eq:mc1}), $i$ is a p-Null player in $(N, u_T, p^N)$. Thus, using ENP, 
$$\Phi_{i}(N, u_{T},p^{N})=0,\;\;\textrm{for all $i \notin T$}.$$
Now by EE for the probabilistic TU game $(N, u_T, p^N)$, 
\begin{eqnarray*}
\mathbb{E}(N,u_{T},p^{N})&=&\sum_{i\in N(p^{N})}\Phi_{i}(N, u_{T},p^{N})\\
&=&\sum_{i\in T}\Phi_{i}(N, u_{T},p^{N}).
\end{eqnarray*}
Since $\Phi_{i}(N, u_{T},p^{N})=0$, for all $i \notin T$ therefore, using COM we have,\\
 $\Phi_{i}(N, u_{T},p^{N})=\left\lbrace \begin{array}{cc}
\frac{\mathbb{E}(N, u_{T},p^{N})}{|T|} & \forall~~i \in T.\\
0 & \text{otherwise}
\end{array}\right.\\
~~~~~~~~~~~~~~~~~~~=\Phi^{Exp-Sh}_{i}(N, u_{T}, p^{N})$.\\
This completes the proof.
\end{proof}
{\begin{remark}
The logical independence of each of the axioms in theorem \eqref{them:ESLN} are shown below: 
\begin{enumerate}[~(a)]
\item $\Phi_{i}(N,v,p^{N})=\frac{\mathbb{E}(N,v,p^{N})-\mathbb{E}(N,v,p^{N}_{-i})}{|N|}$, for all $i \in N$ satisfies ENP, COM, ADD but not the EE.
\item $\Phi_{i}(N,v,p^{N})=\frac{\mathbb{E}(N,v,p^{N})}{|N|}$, for all $i \in N$ satisfies EE , COM, ADD but not  ENP.
  \item If $\Phi_{i}(N,v,p^{N})=\frac{\mathbb{E}(N,v,p^{N})-\mathbb{E}(N,v,p^{N}_{-i})\cdot \mathbb{E}(N,v,p^{N})}{|N|\sum_{i \in N}\mathbb{E}(N,v,p^{N})-\mathbb{E}(N,v,p^{N}_{-i})}$, then for an arbitrary $k \in \mathbb{R}$, define\\
   \begin{equation*}\Phi_{i}(N,v,p^{N})=\left\lbrace \begin{array}{cc}
   \Phi_{i}^{Exp-Sh}(N,v,p^{N})& v(N) \leq k\\
 \Phi_{i}(N,v,p^{N})   &  v(N)> k
    \end{array}\right.
    \end{equation*}
Then $\Phi$ satisfies EE, COM,  ENP  but not ADD.
    \item  For $N \in \aleph $ and $(N, v, p^N)\in G \times \mathbb P$, let $\overline{n}(N)$ be the lowest leveled player in $N$ i.e., $v(i) \geq v(\overline{n}(N))$ for all $i \in N$. Then the value $\Phi$ on $G \times \mathbb P$ given by,\\
     $\Phi_{i}(N	,v,p^{N})=\left\lbrace \begin{array}{cc}
 0 & i = \overline{n}(N)\\
 \dfrac{X_{i}}{\displaystyle\sum_{i \in N \setminus \overline{n}(N)} X_{i}}\mathbb{E}(N,v,p^{N}) & i \in N \setminus \overline{n}(N)
 \end{array}\right.$\\ where $X_{i}=\mathbb{E}(N,v,p^{N})-\mathbb{E}(N,v,p^{N}_{-i})$ satisfies EE, ENP, ADD but not  COM.
\end{enumerate}
\end{remark}}
\subsection{The Second Characterization}\label{subsec:4.2}
\noindent In \cite{laruelle}, the Balanced Contribution property is introduced for values on the class of probabilistic TU games as an extension of BC for classical TU games due to \cite{myerson}. We use this axiom but call it the Expected Balanced contribution (EBC) to distinguish it from the original BC for classical TU games. We show that the EBC together with EE characterizes the Expected Shapley value. Recall that for the class of classical TU games, a  value $\Phi$ satisfies the BC if for every pair of players $i,j\in N$, $\Phi_{i}(N,v)-\Phi_{i}(N\setminus j,v)=\Phi_{j}(N,v)-\Phi_{j}(N\setminus i,v)$, where $(N\setminus i,v)$ is a restriction of $v$ on $N\setminus i \;\;\textrm{for all $S \subseteq N \setminus i$}$. In a similar manner, we have the following.
\begin{definition}
A value $\Phi$ on $G \times \mathbb P$ satisfies Expected Balanced Contribution (EBC) if for any $(N, v,p^{N}) \in G \times \mathbb{P}$, $N\in \aleph$ and for all $i,j \in N,~~i\neq j$, we have, 
$$\Phi_{i}(N, v,p^{N})-\Phi_{i}(N, v,p^{N}_{-j})=\Phi_{j}(N, v, p^{N})-\Phi_{j}(N, v, p^{N}_{-i}).$$
\end{definition}
\begin{theorem}\label{them:bc}
A value $\Phi$ on $G\times \mathbb P$ satisfies EE and EBC if and only if  $\Phi \equiv \Phi^{Exp-Sh}$.
\end{theorem}
\begin{proof}
We have already shown that the Expected Shapley value satisfies EE. Observe that, for any $i\in N$,
\begin{eqnarray*}
N(p^N_{-i})&=& \Big\{S\subseteq N:p^{N}_{-i}(S)>0\Big\}\\
&=&\Big\{S\subseteq N: p^{N}(S)>0\Big\}\setminus \Big\{S \subseteq N: i \in S \Big\}\\
&=& N(p^{N})\setminus \Big\{S \subseteq N: i \in S \Big\}
\end{eqnarray*}
It follows from the above that the inclusion $S \subseteq N(p^N_{-i})$ is equivalent to  $S \in N(p^{N})\setminus \{i\}$.
Let $(N, v, p^N) \in G \times \mathbb P$, $i,j\in N$ and $i \ne j$. Then, using Eq.(\ref{eq:p_i}) we have,
\begin{eqnarray*}
\Phi_{i}^{Exp-Sh}(N, v,p^{N})-\Phi_{i}^{Exp-Sh}(N, v,p^{N}_{-j})~~~~~~~~~~~~~~~~~~~~~~~~~~~~~~~~~~~~\\
= \displaystyle\sum_{S \subseteq N(p^{N})}p^{N}(S)\Phi^{Sh}_{i}(S,v)-\sum_{S \subseteq N(p^{N}_{-j})}p^{N}_{-j}(S)\Phi^{Sh}_{i}(S,v)\\
= \displaystyle\sum_{{S \in N(p^{N})}: j \notin S, T \subseteq j}p^{N}(S \cup T)\Phi^{Sh}_{i}(S \cup T,v)~~~~~~~~~~~~~\\
-\displaystyle\sum_{{S \in N(p^{N})}: j \notin S}p^{N}_{-j}(S)\Phi^{Sh}_{i}(S,v)~~~~~~~~~~~~~~~~~\\
= \displaystyle\sum_{{S \in N(p^{N}): j \notin S, T \subseteq \{j\}}}p^{N}(S \cup T)\Phi^{Sh}_{i}(S \cup T,v)~~~~~~~~~~~~~\\
-\displaystyle\sum_{{S \in N(p^{N})} \setminus \{j\}}\Big \{p^{N}(S)+p^{N}(S \cup j)\Big \}\Phi^{Sh}_{i}(S,v)\\
=\displaystyle \sum_{S \subseteq N(p^{N}): j \notin S, T \subseteq \{j\}}p^{N}(S \cup T)\Phi^{Sh}_{i}(S \cup T,v)~~~~~~~~~~~~~\\
-\displaystyle\sum_{{S \in N(p^{N}): j \notin S}}p^{N}(S)\Phi^{Sh}_{i}(S,v)~~~~~~~~~~~~~~~~~~\\
-\sum_{{S \in N(p^{N}): j \notin S}}p^{N}(S \cup j)\Phi^{Sh}_{i}(S,v)~~~~~~~~~~~~\\
= \displaystyle \sum_{{S \in N(p^{N}): j \notin S}}p^{N}(S \cup j)\Big \{\Phi^{Sh}_{i}(S \cup j,v)-\Phi^{Sh}_{i}(S,v)\Big \}.~~
\end{eqnarray*}
\noindent For ${S \in N(p^{N})}:j\in S,~~ \Phi^{Sh}_{i}(S \cup j,v)-\Phi^{Sh}_{i}(S,v)=0$, therefore these expressions can be added to the expression derived above. Thus, we get the following.
\begin{equation}\label{EQ:1}\Phi_{i}^{Exp-Sh}(N, v,p^{N})-\Phi_{i}^{Exp-Sh}(N, v,p^{N}_{-j})=\sum_{{S \in N(p^{N})}:i\in S}p^{N}(S)\Big\{\Phi^{Sh}_{i}(S,v)-\Phi^{Sh}_{i}(S\setminus j,v)\Big\}
\end{equation}
 Similarly,
\begin{equation}\label{EQ:2}
\Phi_{j}^{Exp-Sh}(N, v,p^{N})-\Phi_{j}^{Exp-Sh}(N, v,p^{N}_{-i})=\sum_{{S \in N(p^{N})}:j \in S}p^{N}(S)\Big\{\Phi^{Sh}_{j}(S,v)-\Phi^{Sh}_{j}(S\setminus i,v)\Big\}
\end{equation}
From BC of the Shapley value for classical TU games, the required result follows immediately.\\
Next, we show that there exists exactly one value on $G \times \mathbb P$ satisfying EE and EBC.\\
Let $\Phi^{1}$ and $\Phi^{2}$ be two value on $G \times \mathbb P$. 
For $N=\{1\}\in \aleph$, the results hold trivially due to EE of the Shapley value i.e., 
$\Phi_{1}^{1}(\{1\}, v,p^{\{1\}})=p^{\{1\}}(\{1\})v(\{1\})=\Phi_{1}^{2}(\{1\},v,p^{\{1\}})$. Let us assume that for $N_{0}=\{1,2,\ldots,n_{0}\}$ also, the results hold, i.e., for any two arbitrary values $\Phi^{1}$ and $\Phi^{2}$ on $G \times \mathbb P$ satisfying EE and EBC we have, $$\Phi^{1}(N_0, v,p^{N_{0}})=\Phi^{2}(N_0, v,p^{N_{0}})=\Phi^{Exp-Sh}(N_0, v,p^{N_{0}}).$$ 
Let $(N,v,p^{N})\in G\times \mathbb{P}$ such that $N$ has now $|N_{0}|+1$ players. Thus, $N(p_{-i}^{N}) \leq |N_{0}|$ and $N(p_{-j}^{N}) \leq |N_{0}|$.  By induction hypothesis, we have
$$\Phi^{1}(N \setminus i, v,p_{-i}^{N})=\Phi^{2}(N \setminus i, v,p_{-i}^{N})\;\textrm{and}\;\;\Phi^{1}(N \setminus j,v,p_{-j}^{N})=\Phi^{2}(N \setminus j,v,p_{-j}^{N}).$$
By EBC and the induction hypothesis on $\Phi^{1}$ and $\Phi^{2}$ we have,
\begin{eqnarray*}
\Phi_{i}^{1}(N,v,p^{N})-\Phi_{j}^{1}(N,v,p^{N})&=&\Phi_{i}^{1}(N,v,p_{-j}^{N})-\Phi_{j}^{1}(N,v,p_{-i}^{N})\\
&=&\Phi_{i}^{2}(N, v,p_{-j}^{N})-\Phi_{j}^{2}(N,v,p_{-i}^{N}).
\end{eqnarray*} 
It follows that, there exist  $\xi \in \mathbb{R}: \forall i,j \in N$,
$$\Phi_{i}^{1}(N, v,p^{N})-\Phi_{i}^{2}(N,v,p^{N})=\Phi_{j}^{1}(N,v,p^{N})-\Phi_{j}^{2}(N,v,p^{N})=\xi.$$
 By EE, we have $\mathbb{E}(N,v,p^{N}) = \displaystyle\sum_{i\in N}\Phi_{i}^{1}(N,v,p^{N})=\sum_{i\in N}\Phi_{i}^{2}(N,v,p^{N})$. \\
This further implies,
\begin{equation*}
\sum_{i\in N}\Phi_{i}^{1}(N,v,p^{N})-\sum_{i\in N}\Phi_{i}^{2}(N,v,p^{N})= \sum_{i\in N}\Big\{\Phi_{i}^{1}(N,v,p^{N})-\Phi_{i}^{2}(N,v,p^{N})\Big\}= |N| \cdot \xi =0.
\end{equation*}
Therefore, we have $\xi=0$. It follows that $\Phi_{i}^{1}(N, v,p^{N})=\Phi_{i}^{2}(N, v,p^{N})$, for all $i \in N$. 
Hence the uniqueness follows.
\end{proof}
\begin{remark}
\noindent The logical independence of the axioms in theorem \ref{them:bc} are shown below: 
\begin{enumerate} 
\item The value $\Phi_{i}(N,v,p^{N})=\frac{\mathbb{E}(N,v,p^{N})}{{|N|}}$, satisfies EE but not EBC.
\item The value $\Phi_{i}(N,v,p^{N})=\Phi^{Exp-Sh}_{i}(N,v,p^{N})+k,~k \in \mathbb{R}$ satisfies  EBC but not EE. 
\end{enumerate}
\end{remark}
\begin{remark}
Note that, similar to BC on classical TU games, we can justify that EBC is a reasonable property for any value to satisfy under the probabilistic set up since it brings about some kind of stability to the coalition formation process. It is interesting to note further that the expected marginal contribution due to \cite{laruelle} satisfies EBC only for a special class of probability distributions $p^N \in \mathbb P$ which satisfies the condition $\displaystyle\sum_{T: i\in T} p^N (T) =  \sum_{T: j\in T} p^N (T)$ for any pair of players $i,j\in N$. On the contrary, there is no such restriction on the probability distribution for the Expected Shapley value.
\end{remark}
\subsection{Characterization in terms of potential:}\label{subsec:4.3} 
In this section, we provide another characterization of the Expected Shapley value in terms of a potential function following the work of \cite{Hart} as mentioned in section~\ref{sec:2}. We define the probabilistic potential function and show that the marginal contribution of each player with respect to this potential function is the Expected Shapley value. We also define two axioms namely, p-Consistency (p-CON) and Standard for two person probabilistic games (STPPG) similar to CON and STPG of classical TU games.
\begin{definition}
Given a function $\P:G\times \mathbb{P} \to \mathbb{R}$, the probabilistic marginal contribution of player $i \in N$ in a probabilistic game $(N, v, p^N)$ with respect to $\P$ is given by 
\begin{equation}\label{eq:probpot}
D_{i}\P(N,v,p^{N})=\P(N, v,p^{N})-\P(N\setminus i, v, p_{-i}^{N\setminus i}),
\end{equation}
where $(N \setminus i, v, p^{N\setminus i}_{-i})$ is the restriction of $(N, v, p^N)$ with $\P(\emptyset, v, p^\emptyset_\emptyset)=0$. Note that, now we abuse the notations to denote by 
$p^T_T$ the restriction of $p^N_T$ to $T$ and in particular by $p^{N\setminus i}_{-i}$  the restriction of $p^N_{-i}$ to $N \setminus i$.\\
A function $\P:G\times \mathbb{P} \to \mathbb{R}$ is called a probabilistic potential function if it satisfies the following condition.
\begin{equation}\label{potential:1}
\sum_{i \in N}D_{i}\P(N, v, p^{N})=\mathbb{E}(N, v, p^{N})\;\;\textrm{for all $(N, v, p^N)\in G\times \mathbb P$}
\end{equation}
\end{definition}
\begin{remark}
It follows from Eq.(\ref{potential:1}) that
\begin{eqnarray}
\sum_{i \in N}D_{i}\P(N, v,p^{N})=\mathbb{E}(N, v,p^{N})~~~~~~~~~~~~~~~~~~~~~~~~~~~~~~~~~~~~~~~~~~~~~~~~~~~~~~~~~~~~~~~~~~~~~~~~~~~\nonumber\\
\Rightarrow  \sum_{i \in N}\{\P(N,v,p^{N})-\P(N\setminus i, v,p^{N\setminus i}_{-i})\}=\mathbb{E}(N,v,p^{N})~~~~~~~~~~~~~~~~~~~~~~~~~~~~~~~~~~\nonumber\\
\Rightarrow  |N|\P(N, v, p^{N})- \sum_{i \in N} \P(N \setminus i,v,p^{N\setminus i}_{-i})=\mathbb{E}(N,v,p^{N})~~~~~~~~~~~~~~~~~~~~~~~~~~~~~~~~\nonumber\\
\Rightarrow  \P(N,v,p^{N})=\frac{1}{|N|}\Big\{\mathbb{E}(N,v,p^{N})+\sum_{i \in N}\P(N\setminus i, v,p_{-i}^{N\setminus i})\Big\}.~~~~~~~~~~~~~~~~~~~~~~~~~~\label{eq:pot11}
\end{eqnarray}
\end{remark}
\begin{remark}
Note that the notion of a potential function is also discussed in \cite{laruelle}, where they consider the potential of a probabilistic TU game as the expected worth of the coalition that would form. However, in our model, we define the potential function recursively in line with \cite{Hart} such that the probabilistic marginal contributions, one for each player add up to the expected worth of the game. Recall from section~\ref{sec:2}~ that this is similar to the potential function of the classical TU game due to \cite{Hart} where the marginal contributions (Shapley payoffs to the players) add up to the worth of the grand coalition. In what follows next, we show that the probabilistic marginal contributions given by Eq.(\ref{eq:probpot}) are eventually the payoffs to the players under the probabilistic Shapley value. Thus, under the probabilistic setup, we do not deviate much from the original narratives of the potential function of classical TU games.
\end{remark}

\begin{proposition}
There exists a unique probabilistic potential function $\P$. For each $(N,v,p^{N}) \in G\times \mathbb{P}$, we must have $D_{i}\P(N,v,p^{N})=\Phi_{i}^{Exp-Sh}(N,v,p^{N})$. The probabilistic potential of any probabilistic TU game is determined uniquely by Eq.(\ref{potential:1}).
\end{proposition}
\begin{proof}
First we prove the existence of a probabilistic potential function. \\
Define, 
\begin{equation}\label{eq:pot1}
\P(N,v,p^{N})=\displaystyle\sum_{{S \in N(p^{N})}}p^{N}(S)P(S,v),
\end{equation}
where $P:G \to \mathbb{R}$ is the unique potential function on the class of classical TU games. Now, 
\begin{eqnarray}
{\sum_{i \in N}D_{i}\P(N, v,p^{N})}&=&\sum_{i \in N}\Big\{\P(N,v,p^{N})-\P(N\setminus i, v,p_{-i}^{N\setminus i}) \Big\}\nonumber\\
&=& \sum_{i \in N} \Big\{ \sum_{S \subseteq N} p^{N}(S)P(S,v)-\sum_{{S \in N(p_{-i}^{N\setminus i})}} p_{-i}^{N\setminus i}(S)P(S,v)\Big\}\nonumber\\
&=& \sum_{i \in N} \Big\{\sum_{S \subseteq N(p^{N}):i\in S} p^{N}(S)P(S,v)-\Big(\sum_{{S \in N(p_{-i}^{N})}}P(S,v)p_{-i}^{N}(S)\nonumber\\
&& ~~~~~~~~~~+\sum_{{S \in N(p^{N}): i \notin S}}p^{N}(S)P(S,v)\Big)\Big\} \nonumber\\
&= &\sum_{i \in N} \Big\{\sum_{S \subseteq N(p^{N}):i\in S} p^{N}(S)P(S,v)\nonumber\\
&&~~~~~~~~~~~+\sum_{{S \in N(p_{-i}^{N}):i\in S}}P(S,v)\big[p^{N}(S)-p^{N}_{-i}(S)\big]\Big\}\nonumber\\
&= & \sum_{i \in N} \Big\{\sum_{{S \in N(p^{N}):i\in S}} p^{N}(S)P(S,v)\nonumber\\
&& ~~~~~~~~~~~-\sum_{S \subseteq N(p_{-i}^{N}):i\in S} p^{N}(S \cup i)P(S,v) \Big\}\nonumber\\
&=& \sum_{i \in N} \Big\{\sum_{{S \in N(p^{N}):i\in S}} p^{N}(S)\big\{P(S,v)-P(S \setminus i,v)\big\}\Big\}\nonumber\\
&=& \sum_{i \in N} \Big\{\sum_{{S \in N(p^{N}):i \in S}} p^{N}(S)\Phi_{i}^{Sh}(S,v)\Big\}\label{eq:pot2}
\end{eqnarray}
It follows from the derivation of Eq.(\ref{potential:1}) given by Eq.(\ref{eq:pot11}), that 
starting from $\P(\emptyset,v,p^{N}_{\emptyset})=0$, we get a uniquely determined probabilistic potential function.\\
Moreover, from Eq.(\ref{eq:pot2}) and the definition of the Expected Shapley value, we have 
\begin{equation*}
\sum_{i \in N}D_{i}\P(N, v,p^{N}) = \sum_{i \in N}\Phi_{i}^{Exp-Sh}(N, v, p^N)
\end{equation*}
By EBC of the Expected Shapley value, $D_{i}\P(N, v,p^{N})=\Phi_{i}^{Exp-Sh}(N, v, p^N)$ for all $i \in N$. This completes the proof.
\end{proof}
\noindent Next, we introduce the two axioms that characterize the Expected Shapley value mentioned in the beginning of this section. However, before that, we define a probabilistic reduced game with respect to a value on $G$.  Then, as a particular case, we define the probabilistic reduced game with respect to the Shapley value. As mentioned in section~\ref{sec:2}, we use the notion of value dividend introduced in Eq.(\ref{eq:value-dividend}), to define the reduced game in the probabilistic framework and subsequently to characterize the Expected Shapley value using the notion of the potential of a game.
\begin{definition}\label{def:prob_reduced-game}
The probabilistic reduced game of a probabilistic TU game $(N, v, p^N)$ on $T \subseteq N$ with respect to an arbitrary value $\Phi$  is the {triple $(T,v^{^{p^{N}}}_{T,\Phi}, p_T^{T})$ where $v^{^{p^{N}}}_{T,\Phi}$ is given by the following expression:}
\begin{eqnarray}\label{eq:prob_red}
{v_{T,\Phi}^{^{p^{N}}}(S)}
&=&\sum_{  K_{1} \subseteq S,  K_{2} \subseteq T^{c}: K_{1} \neq \emptyset}\thetak {K_{1} \cup K_{2}}\frac{p^{N}_{K_{1} \cup K_{2}}(K_{1} \cup K_{2})}{p^{N}_{K_{1}}(K_{1})},~~\forall ~ S \subseteq T.
\end{eqnarray}
 and ${v_{T,\Phi}^{^{p^{N}}}}(\emptyset)=0$ and $p^T_T$ is the restriction of $p^N_T$ to $T$.
\end{definition}
\begin{remark}
Recall from section~\ref{sec:2} that the reduced game on a coalition $T$ in classical TU games is defined by rescaling the worth of the coalitions so that the players outside $T$ get their payoffs according to this value and leave the game. This rescaling of the worth in the reduced game given by Eq.(\ref{eq:vdreduced}) in value dividend form is determined by the value dividend $\Theta^\phi_v$. Comparing the two equations Eq.(\ref{eq:vdredSh_1}) and Eq.(\ref{eq:prob_red_Sh_1}) we can see that they differ by the probability term $\frac{p^{N}_{(K_{1} \cup K_{2})}(K_{1} \cup K_{2})}{p^{N}_{K_{1}}(K_{1})}$. This we interpret as the conditional probability that the players outside $T$ get their payoffs according to the value $\Phi$ and leave the game subject to the realization of all the coalitions of $T$. This is done with a rescaling of the probabilities determined by the original probability distribution $p^N$ restricted to the coalitions of $T$.
\end{remark}
\noindent In particular, the probabilistic reduced game {$(T, v^{^{p^{N}}}_{T,\Phi}, p_T^T)$} with respect to the  Shapley value in value dividend form can be obtained from Eq.(\ref{eq:prob_red}) and using similar arguments as in Eq.(\ref{eq:vdredSh_1}) as follows:
{\begin{eqnarray}\label{eq:prob_red_Sh_1}
v_{T,\Phi}^{^{p^{N}}}(S)&=&\sum_{K_{1} \subseteq S,K_{2} \subseteq T^{c}:K_{1} \neq \emptyset}|K_{1}|\cdot
{\frac{\Theta_{v}^{\Phi^{Sh}}(K_{1} \cup K_{2})}{|K_{1}|+|K_{2}|}}\frac{p^{N}_{(K_{1} \cup K_{2})}(K_{1} \cup K_{2})}{p^{N}_{K_{1}}(K_{1})},~~\forall ~ S \subseteq T.\\
&=&\displaystyle \sum_{K_{1} \subseteq S,K_{2} \subseteq T^{c}:K_{1} \neq \emptyset}|K_{1}|\cdot
{\frac{\Delta_{v}(K_{1} \cup K_{2})}{|K_{1}|+|K_{2}|}}\frac{p^{N}_{(K_{1} \cup K_{2})}(K_{1} \cup K_{2})}{p^{N}_{K_{1}}(K_{1})},~~\forall ~ S \subseteq T.\\
\end{eqnarray}}
\noindent The corresponding axioms for characterization of the Expected Shapley value proceed as follows.
\begin{axiom}[p-Consistency (p-CON)]
A value $\Phi$ on $G \times \mathbb P$ satisfies p-Consistency (p-CON) namely, 
\begin{equation}\label{eq:pcon}
\Phi_i(N, v, p^N) = {\Phi_i(T, v_{T,{\Phi}}^{p^N}, p_T^T)\;\;~~\forall~ i\in T}.
\end{equation} 
\end{axiom}
\begin{axiom}[Standard for two person probabilistic games(STPPG)]
A value $\Phi$ on $G \times \mathbb P$  satisfies Standard for two person probabilistic games namely, for $N=\{i,j\}$ and a two person probabilistic game $(N, v, p^N)$,
\begin{equation}\label{eq:stppg}
\Phi_i(\{i,j\}, v, p^N) = \mathbb E(\{i,j\}, v, p^N_{-j}) + \frac{1}{2} p^N(\{i,j\})\Big\{v(\{i,j\})-v(i)-v(j)\Big\}.
\end{equation} 
\end{axiom}
\begin{proposition}
The Expected Shapley value satisfies p-CON and STPPG.
\end{proposition}
\begin{proof}
\noindent The Expected Shapley value satisfies p-CON:
From Eq.(\eqref{eq:vdreduced}), we have:
\begin{equation}\label{reduced-game-dividned}
{\frac{\Delta_{v_{T,\Phi}^{p^{N}}}(K_{1})}{|K_{1}|}}
=\sum_{ K_{2} \subseteq T^{c}} \frac{\Delta_{v}( K_{1} \cup K_{2})}{|K_{1}|+|K_{2}|}\Big\{\frac{p^{N}_{K_{1} \cup K_{2}}(K_{1} \cup K_{2})}{p^{N}_{K_{1}}(K_{1})}\Big\},~~\forall ~K_{1} \subseteq T.
\end{equation}
It follows from Eq.(\ref{reduced-game-dividned}),
\begin{equation*}
\sum_{K_{1}\subseteq S: i \in K_{1}}\frac{\Delta_{v_{T,\Phi}^{p^{N}}}(K_{1})}{|K_{1}|}
=\sum_{K_{1}\subseteq S: i \in K_{1}}\sum_{ K_{2} \subseteq T^{c}} \frac{\Delta_{v}(K_{1} \cup K_{2})}{|K_{1}|+|K_{2}|}\Big\{\frac{p^{N}_{K_{1} \cup K_{2}}(K_{1} \cup K_{2})}{p^{N}_{K_{1}}(K_{1})}\Big\},~~\forall ~K_{1} \subseteq T.
\end{equation*}
Therefore, we have
\begin{equation*}
{\Phi_{i}^{Sh}(S,v_{T,\Phi}^{p^{N}})}
 =\sum_{\substack{K_{1} \subseteq S: i \in K_{1}\\ K_{2} \subseteq T^{c}}}\frac{\Delta_{v}( K_{1} \cup K_{2})}{|K_{1}|+|K_{2}|}\cdot \frac{\Big\{p^{N}_{K_{1} \cup K_{2}}(K_{1} \cup K_{2})\Big\}}{p^{N}_{K_{1}}(K_{1})},~~\forall ~S\subseteq T~~~~~~~~~~~~~~~~~~~~~~~~~~~~~~~~~~
\end{equation*}
It follows that,
\begin{eqnarray*}
\Phi_{i}^{Exp-Sh}(T,v_{T,\Phi}^{p^{N}},p^{T}_{T})&=&\sum_{S \subseteq T:i \in S}p^{N}_{T}(S)\Phi_{i}^{Sh}(S,v_{T,\Phi}^{p^{N}}) \\
& = & \sum_{S \subseteq T:i \in S}p^{N}_{T}(S)\sum_{i \in K_{1}: K_{1} \subseteq S,K_{2} \subseteq T^{c}}\frac{\Delta_{v}( K_{1} \cup K_{2})}{|K_{1}|+|K_{2}|}\cdot \frac{\Big\{p^{N}_{K_{1} \cup K_{2}}(K_{1} \cup K_{2})\Big \}}{p^{N}_{K_{1}}(K_{1})}\\
&=&\sum_{i \in K_{1}: K_{1} \subseteq S,K_{2} \subseteq T^{c}}\frac{\Delta_{v}( K_{1} \cup K_{2})}{|K_{1}|+|K_{2}|}\cdot \Big\{p^{N}_{K_{1} \cup K_{2}}(K_{1} \cup K_{2})\Big\}\\
&=&  \sum_{S \subseteq N:i \in S} \frac{\Delta_{v}(S)}{|S|}\Big\{\sum_{K \subseteq N \setminus S}p^{N}_{S}(K)\Big\}\\
&=&  \sum_{S \subseteq N:i \in S} \frac{\Delta_{v}(S)}{|S|}p^{N}_{S}(S)\\
&=& \Phi_{i}^{Exp-Sh}(N,v,p^{N})
\end{eqnarray*}
\noindent The Expected Shapley value satisfies STPPG:
For $N=\{i,j\}$.
\begin{eqnarray*}
\Phi_{i}^{Exp-Sh}(\{i,j\},v,p^{N})&=& \sum_{S \subseteq N: i \in S}p^{N}(S)\Phi_{i}^{Sh}(S,v)\\
&=&p^{N}(i)\Phi_{i}^{Sh}(\{i\},v)+p^{N}(i,j)\Phi_{i}^{Sh}(\{i,j\},v)\\
&=&p^{N}(i)v(i)+p^{N}(i,j)\Big\{\frac{\Delta_{v}(\{i,j\})}{2}+v(i)\Big\}\\
&=& \mathbb{E}(\{i,j\},v,p^{N}_{-j})+ p^{N}(i,j)\Big\{\frac{\Delta_{v}(\{i,j\})}{2}\Big\}\\
&=& \mathbb{E}(\{i,j\},v,p^{N}_{-j})+ p^{N}(i,j)\Big\{\frac{v(\{i,j\})-v(i)-v(j)}{2}\Big\}
\end{eqnarray*}
\end{proof}
\begin{theorem}\rm \label{them:Potential}
A value $\Phi$ on $G \times \mathbb P$ satisfies STTPG and p-CON if and only if, $\Phi(N,v,p^{N})\equiv \Phi^{Exp-Sh}(N,v,p^{N})$ for all $(N, v, p^N) \in G \times \mathbb P$.
\end{theorem}
\begin{proof}
It is already shown that the Expected Shapley value satisfies p-CON and STPPG. Conversly we prove that any value satisfying  p-CON and STPPG   also satisfy Expected  Efficiency. For $N=\{i,j\}$,   p-CON and STPPG  implies Expected  Efficiency. Consider the game $(v,p^{\{i\}})$, we define the two player game $v^{*}$ and probability distribution $p^{\{i,j\}}$ such that, $p^{\{i,j\}\setminus j}_{-j}=p^{\{i\}}$ and $v^{*}(i)=v(i),~v^{*}(j)=0,~v^{*}(i,j)=v(i)$. By definition of reduced game 
$v^{{*}^{p^{N}}}_{i}(i)=v(i)$. Since $\Phi$ satisfies  p-CON therefore we have,\\ 
$\Phi_{i}(\{i\},v,p^{\{i\}})=\Phi_{i}(\{i\},v^{{*}^{p^{N}}}_{i},p^{\{i,j\}\setminus j}_{-j})= \Phi_{i}(\{i,j\},v^{*},p^{\{i,j\}})=v^{{*}^{p^{N}}}_{i}(i)p^{\{i,j\}\setminus j}_{-j}(i)=v(i)p^{\{i\}}(i)$.
Next, for $|N| \geq 3$, suppose that for all games  with player set $T$ such that $|T|<|N|$, a value on $G \times \mathbb P$ satisfying p-CON and STTPG also satisfies EE. Then,
\begin{eqnarray*}
\sum_{i \in N}\Phi_{i}(N,v,p^{N})&=& \Phi_{k}(N,v,p^{N}) + \sum_{i \in  N \setminus k}\Phi_{i}(N,v,p^{N})\\
&=& \Phi_{k}(N,v,p^{N}) + \sum_{i \in  N \setminus k}\Phi_{i}(N \setminus k,v_{N \setminus k,\Phi}^{p^{N}},p^{N \setminus k}_{-k})\\
&=& \Phi_{k}(N,v,p^{N})+\mathbb{E}(N \setminus k,v_{N \setminus k,\Phi}^{p^{N}},p^{N \setminus k}_{-k})\\
&=& \mathbb{E}(N,v,p^{N})
\end{eqnarray*}
Hence by induction hypothesis we conclude that $\Phi$ is an efficient value function.\\
Next if a Potential function $\Q:G \times \mathbb{P} \to \mathbb{R}$  can be constructed such that  $\Q(\emptyset,v,p^{\emptyset}_{\emptyset})=0$ and 
\begin{equation}\label{def:Q}
\Phi_{i}(N,v,p^{N})= \Q(N,v,p^{N})-\Q(N \setminus i,v,p^{N\setminus i}_{-i}),~ \forall~ i \in N
\end{equation} then by uniqueness of Potential function it follows that $\Q=\P$ and hence $\Phi_{i}(N,v,p^{N})=\Phi_{i}^{Exp-Sh}(N,v,p^{N})$. 
Therefore, the existence of such a $\Q$ completes the proof.\\ 
Define, $\Q(\emptyset,v,p^{\emptyset}_{\emptyset})=0,~\Q(\{i\},v, p^{i}_{i})=p^{i}_{i}(i)v(i)$, $\Q(\{i,j\},v, p^{\{i,j\}}_{\{i,j\}})=p^{\{i,j\}}_{\{i,j\}}(i,j)\Big\{\frac{\Delta_{v}(\{i,j\})}{2}+\Delta_{v}(i)+\Delta_{v}(j)\Big\}+p^{\{i,j\}}_{\{i,j\}}(i)\Delta_{v}(i)+p^{\{i,j\}}_{\{i,j\}}(j)\Delta_{v}(j)$. Let $|N| \geq 3$ and $\Q$ is already defined as \eqref{def:Q} for all coalitions of size less then $|N|$. Let us define $\Q(N,v,p^{N})=\alpha$ if and only if $\alpha-\Q(N \setminus i,v,p^{N \setminus i}_{-i})=\Phi_{i}(N,v,p^{N})$ for all $i \in N$. Next, we show that, 
\begin{equation}\label{unique-relation:Q}
\Q(N \setminus i,v,p^{N \setminus i}_{-i})+\Phi_{i}(N,v,p^{N})=\Q(N \setminus j, v,p^{N \setminus j}_{-j})+\Phi_{j}(N, v,p^{N}),~~\forall ~i,j \in N.
\end{equation}
Let $|K| \in N \setminus \{i,j\}, |N| \geq 3$.
\begin{eqnarray*}
&&\Phi_{i}(N,v,p^{N})-\Phi_{j}(N,v,p^{N})\\
&=& \Phi_{i}(N \setminus k, v^{p^{N}}_{N \setminus k},p^{N \setminus k}_{-k})-\Phi_{j}(N \setminus k,v^{p^{N}}_{N \setminus k},p^{N \setminus k}_{-k})\\
&=& Q(N \setminus k,v_{N \setminus k,\Phi}^{p^{N}},p^{N \setminus k}_{-k})-Q(N\setminus \{k,i\},v_{N\setminus \{k,i\},\Phi}^{p^{N}},p^{N\setminus \{k,i\}}_{-\{k,i\}})\\
&&~-Q(N \setminus k,v_{N \setminus k,\Phi}^{p^{N}},p^{N \setminus k}_{-k})+Q(N\setminus \{k,j\},v_{N\setminus \{k,j\},\Phi}^{p^{N}},p^{N\setminus \{k,j\}}_{-\{k,j\}})\\
&=&-Q(N\setminus \{k,i\},v_{N\setminus \{k,i\},\Phi}^{p^{N}},p^{N\setminus \{k,i\}}_{-\{k,i\}})+Q(N\setminus \{i,k,j\},v_{N\setminus \{i,k,j\},\Phi}^{p^{N}},p^{N\setminus \{i,k,j\}}_{-\{i,k,j\}})\\
&& ~ + Q(N\setminus \{k,j\},v_{N\setminus \{k,j\},\Phi}^{p^{N}},p^{N\setminus \{k,j\}}_{-\{k,j\}})-Q(N\setminus \{i,k,j\},v_{N\setminus \{i,k,j\},\Phi}^{p^{N}},p^{N\setminus \{i,k,j\}}_{-\{i,k,j\}})\\
&=& -\Phi_{j}(N \setminus \{k,i\},v_{N \setminus \{k,i\},\Phi}^{p^{N}},p^{N \setminus \{k,i\}}_{-\{k,i\}})+\Phi_{i}(N \setminus \{k,j\},v_{N \setminus \{k,j\},\Phi}^{p^{N}},p^{N \setminus \{k,j\}}_{-\{k,j\}})\\
&=& -\Phi_{j}(N \setminus \{i\},v_{N \setminus \{i\},\Phi}^{p^{N}},p^{N \setminus i}_{-i})+\Phi_{i}(N \setminus \{j\},v_{N \setminus \{j\}}^{p^{N}},p^{N \setminus j}_{-j})\\
&=& -\Q(N\setminus i,v_{N\setminus i,\Phi}^{p^{N}},p^{N\setminus i}_{-i})+\Q(N\setminus \{i,j\},v_{N\setminus \{i,j\},\Phi}^{p^{N}},p^{N \setminus \{i,j\}}_{-\{i,j\}})\\
&&+\;\Q(N\setminus j,v_{N\setminus j,\Phi}^{p^{N}},p^{N\setminus j}_{-j})-Q(N\setminus \{i,j\},v_{N\setminus \{i,j\},\Phi}^{p^{N}},p^{N\setminus \{i,j\}}_{-\{i,j\}})\\
&=& \Q(N\setminus j,v_{N\setminus j,\Phi}^{p^{N}},p^{N\setminus j}_{-j})-\Q(N\setminus \{i\},v_{N\setminus i,\Phi}^{p^{N}},p^{N\setminus i}_{-i})
\end{eqnarray*}
{The second line of the above equality follows from `reduced game consistency', the third line follows from the `uniqueness of potential'.\\}
Thus, we have
$$\Q(N \setminus i,v,p^{N\setminus i}_{-i})+\Phi_{i}(N,v,p^{N})=\Q(N \setminus j, v, p^{N\setminus j}_{-j})+\Phi_{j}(N,v,p^{N}),~\forall~i ,j \in N.$$
It follows that the function $\Q$ is unique and hence $\Phi_{i}(N,v,p^{N})=\Phi_{i}^{Exp-Sh}(N,v,p^{N})$, for all $i \in N$.
\end{proof}
\begin{remark}
The logical independence of each of the axioms in theorem \eqref{them:Potential} are shown below: 
\begin{enumerate}
    \item $\Phi_{i}(N,v,p^{N})=\left\lbrace \begin{array}{cc}
   \Phi_{i}^{Exp-Sh}(N,v,p^{N})& ~|N|\leq 2\\
 {\frac{v(N)p^{N}(N)}{|N|}} & |N|> 3
    \end{array}\right.$  satisfies STPPG, but not p-CON.
 \item {The value $\Phi_{i}(N,v,p^{N})=k\cdot \Phi_{i}^{Exp-Sh}(N,v,p^{N}),~$ for any $k \in \mathbb{R} \setminus 1$, satisfies p-CON, but not STPPG.}
\end{enumerate}

\end{remark}
\section{Conclusion}\label{sec:5}

We have studied cooperative games in situations where the coalitions are realized with endogeneously given probabilities. The allocation of the expected worth to the players needs to be made before the realization of the state. As a solution concept, we propose a value called Expected Shapley value that allocates the players their expected worth from the Shapley values over all probable coalitions with respect to a probability distribution. We provided three characterizations of this value based on a natural adaptation of the axioms used in the seminal axiomatizations of the Shapley Value. Similar studies may be made with other values for TU games viz., the Equal Division rule, the Egalitarian Shapley value due to \cite{joosten}, the Solidarity value due to \cite{nr} etc., just to name a few. This we keep for our future research.


\begin{thebibliography}{1}



\bibitem{Besner-2020} 
Besner M (2020) Value dividends, the Harsanyi set and extensions, and the proportional Harsanyi solution. International Journal of Game Theory https://doi.org/10.1007/s00182-019-00701-4.

\bibitem{borkotokey_mss}
Borkotokey, S., S. Chakrabarti, R.P. Gilles, L. Gogoi and R. Kumar (2021), Probabilistic Network Values, Mathematical Social Sciences, 113, 169-180

\bibitem{carreras_a}
Carreras F, Puente M (2015a) Multinomial probabilistic values. Group Decis Negot 24(6):981-991 

\bibitem{carreras_b}
Carreras F, Puente M (2015b) Coalitional multinomial probabilistic values. Eur J Oper Res 245(1):236-246.

\bibitem{Chakrabarti-a} 
Chakrabarti, S., Gogoi L.,  Gilles P.R., Borkotokey, S. and Kumar, R.(2021) Expected Values for Variable Network Games.  	https://doi.org/10.48550/arXiv.2108.07047.

\bibitem{calvo}
Calvo, E., J. Lasaga, and A. van den Nouweland (1999) Values of games with probabilistic graphs, Mathematical Social Sciences, 37, 79-95.

\bibitem{ghintan}
Ghintran, A., E. Gonz\'alez-Arangu\"ena, and C. Manuel (2012) A probabilistic position value, Annals of Operations Research, 201, 183-196.

\bibitem{gomez}
G\'omez, D., E. Gonz\'alez-Arangu\"ena, C. Manuel, and G. Owen (2008) A value for generalized probabilistic communication situations, European Journal of Operational Research, 190, 539-556.

\bibitem{harsanyi}
Harsanyi, J.C. (1963) A Simplified Bargaining Model for the $n$ Person Co-operative Game, International Economic Review, 4, 194-220.

\bibitem{Hart}
Hart S,  Mas-Colell A (1989) Potential, value, and consistency. Econometrica: Journal of the Econometric Society, 589-614. 

\bibitem{jw}
Jackson, M. O. and A. Wolinsky (1996) A Strategic Model of Social and Economic Networks, Journal of Economic Theory, 71, 44-74.

\bibitem{joosten}
Joosten, R., (1996) Dynamics, equilibria and values dissertation, Maastricht University.

\bibitem{kamionko}
Kamionko, V., and Marakulin, V. M.,  (2020) Shapley's Value and Its Axiomatization in Games with Prior Probabilities of Coalition Formation. Journal of the New Economic Association, 46, 2020, Available at SSRN: https://ssrn.com/abstract=3524864 or http://dx.doi.org/10.2139/ssrn.3524864.

\bibitem{koster}
Koster, M. · S. Kurz, I. Lindner, and S. Napel (2017) The Prediction value, Social Choice and Welfare, 48, 433-460.


\bibitem{laruelle}
Laruelle, A., F. Valenciano (2008) Potential, value, and coalition formation. TOP 16(1):73-89
\bibitem{Lehrer} Lehrer, E. (1988) An axiomatization of the Banzhaf value, International Journal of Game Theory,
17 , 89-99.

\bibitem{myerson78}
Myerson, R. B. (1977): Graphs and Cooperation in Games, Mathematics of Operations Research, 2, 225-229.

\bibitem{myerson} 
Myerson, R. B. (1980) Conference Structure and Fair allocation rules, International Journal of Game Theory, 9,169-182. 

\bibitem{nr}
 Nowak, A. S., and Radzik, T., (1994) A solidarity value for n-person transferable utility games, International
Journal of Game Theory, 23, 43--48.


\bibitem{pongou}
Pongou, R., and Tondji, J-B. (2017) Valuing inputs under supply uncertainty: The Bayesian Shapley value. Games Econ. Behav., http://dx.doi.org/10.1016/j.geb.2017.08.005.

\bibitem{owen}
Owen, G. (1972) Multilinear extensions of games, Management Sciences 18  64-79.

\bibitem{shapley}
Shapley, L.S. (1953) A value for n-persons games, Annals of Mathematics Studies 28, 307-318.

\bibitem{vladimir}
Kamionko, V. \& Marakulin, V. (2020). Shapley’s value and its axiomatization in games with prior probabilities of coalition formation. Journal of the New Economic Association. 46. 12-29. 

\bibitem{weber}
Weber, R. J., 1988, Probabilistic Values for Games. In Roth, A. E., ed., (1988),
101-119.

\end{thebibliography}
\end{document}